\documentclass[10pt, conference, compsocconf, letterpaper]{IEEEtran}
\IEEEoverridecommandlockouts

\usepackage{cite}
\usepackage{amsmath, amssymb, amsfonts}
\usepackage{algorithm} 
\usepackage{algpseudocode} 
\usepackage{xcolor}
\usepackage{hyperref}
\usepackage{graphicx}
\usepackage{textcomp}
\usepackage{subfig}
\usepackage{multirow}
\usepackage{colortbl}
\usepackage{soul}
\usepackage{enumerate}
\usepackage{graphicx} 
\usepackage{comment}
\definecolor{mintgreen}{rgb}{0.6, 1.0, 0.6}
\definecolor{pastelviolet}{rgb}{0.8, 0.6, 0.79}
\definecolor{peridot}{rgb}{0.9, 0.89, 0.0}
\definecolor{richbrilliantlavender}{rgb}{0.95, 0.65, 1.0}
\definecolor{robineggblue}{rgb}{0.0, 0.8, 0.8}
\usepackage{xcolor,colortbl}
\definecolor{green}{rgb}{0.1,0.1,0.1}

\newcommand{\name}{{\em HeteroEdge }}
\newcommand{\names}{{\em HeteroEdge}}

\def\BibTeX{{\rm B\kern-.05em{\sc i\kern-.025em b}\kern-.08em
    T\kern-.1667em\lower.7ex\hbox{E}\kern-.125emX}}

\begin{document}
\setlength{\abovedisplayskip}{2pt}
\setlength{\belowdisplayskip}{2pt}



\title{HeteroEdge: Addressing Asymmetry in Heterogeneous Collaborative Autonomous Systems}
 \author{\IEEEauthorblockN{Mohammad Saeid Anwar$^1$, Emon Dey$^{1,2}$, Maloy Kumar Devnath$^1$, Indrajeet Ghosh$^{1,2}$, \\ Naima Khan$^{1,2}$, Jade Freeman$^4$, Timothy Gregory$^4$, Niranjan Suri$^4$, Kasthuri Jayarajah$^2$,\\ Sreenivasan Ramasamy Ramamurthy$^3$, Nirmalya Roy$^{1,2}$}
 \IEEEauthorblockA{$^1$Department of Information Systems, University of Maryland Baltimore County (UMBC), $^2$CARDS, UMBC\\
 $^3$Department of Computer Science, Bowie State University, $^4$DEVCOM Army Research Lab\\
 $^1$(saeid.anwar, maloyd1)@umbc.edu, $^{1,2}$(edey1, indrajeetghosh, nkhan4, nroy)@umbc.edu, $^{2}$kasthuri@umbc.edu, \\ $^3$sramamurthy@bowiestate.edu, $^4$(jade.I.freeman2, timothy.c.gregory6, niranjan.suri).civ@army.mil}}

\maketitle

\begin{abstract}
Gathering knowledge about surroundings and generating situational awareness for autonomous systems is of utmost importance for systems developed for smart urban and uncontested environments. For example, a large area surveillance system is typically equipped with multi-modal sensors such as cameras and LIDARs and is required to execute deep learning algorithms for action, face, behavior, and object recognition. However, these systems are subjected to power and memory limitations due to their ubiquitous nature. As a result, optimizing how the sensed data is processed, fed to the deep learning algorithms, and the model inferences are communicated is critical. In this paper, we consider a testbed comprising two Unmanned Ground Vehicle (UGVs)  and two NVIDIA Jetson devices and posit a self-adaptive optimization framework that is capable of navigating the workload of multiple tasks (storage, processing, computation, transmission, inference) collaboratively on multiple heterogenous nodes for multiple tasks simultaneously. The self-adaptive optimization framework involves compressing and masking the input image frames, identifying similar frames, and profiling the devices for various tasks to obtain the boundary conditions for the optimization framework. Finally, we propose and optimize a novel parameter \emph{split-ratio}, which indicates the proportion of the data required to be offloaded to another device while considering the networking bandwidth, busy factor, memory (CPU, GPU, RAM), and power constraints of the devices in the testbed. Our evaluations captured while executing multiple tasks (e.g., PoseNet, SegNet, ImageNet, DetectNet, DepthNet) simultaneously, reveal that executing 70\% (\emph{split-ratio}=70\%) of the data on the \emph{auxiliary node} minimizes the offloading latency by $\approx$ 33\% (18.7 ms/image to 12.5 ms/image) and the total operation time by $\approx$ 47\% (69.32s to 36.43s) compared to the baseline configuration (executing on the \emph{primary node}).
\end{abstract}

\section {\textbf{Introduction}}

In recent days, autonomous systems such as unmanned aerial or ground vehicles have been very popular in various applications, such as surveillance, photography and videography, mapping and surveying, agriculture, environmental monitoring, search and rescue, and delivery services. Unmanned vehicles equipped with sensors such as cameras, lidar, radar, and GPS allow us to collect data and perform tasks (e.g., object detection, scene detection, and many more) in various environments. Performing these tasks while depending on the onboard computational unit (especially advanced deep learning models) involves limited power consumption, eventually affecting the autonomous systems'  operation and safety~\cite{2023mohsanspringer}. 
Several recent studies have investigated the impact of operational time and power supply of autonomous systems~\cite{2022droneflight,2019hashemi, 2021khochare}, and suggest that the operational capacity of the systems is severely affected due to the execution of onboard sub-systems (e.g., navigation unit, cameras, communication systems)~\cite{2021khochare}. In addition to these sub-systems, accommodating Deep Neural Network (DNN)  algorithms (usually power, memory, and computation hungry) to perform the tasks for situational awareness contributes to the already limited power availability~\cite{Heimdel}. Furthermore, as some of the systems' operations, such as navigation and communication, are more important for the safety of the expensive autonomous systems, optimizing the DNNs would be essential to conserve the limited available power.

One of the approaches to conserving power using DNNs can be achieved by offloading the inference task to a remote device (cloud server or a device connected to the same network) with surplus power and computational capability~\cite{xue2022}. However, such a solution is impacted by network availability, reliability, low bandwidth, and latency~\cite{chen2022} caused due to the quality of communication links and the distance between the (\textit{primary node}) (the device that will offload data) and the ~\textit{auxiliary node} (either a remote server or edge device on the same network that can share the workload of the primary node). Some of the limitations of prior research on offloading include an offloading task to homogenous devices (e.g., MASA~\cite{Masa}) and smartphones~\cite{nguyen2020smartphone} and expensive in the case of remote cloud services~\cite{cloud}. For scenarios such as situational awareness by autonomous systems, we hypothesize that leveraging another device within the system would mitigate latency and help conserve power compared to expensive cloud services. Keeping these discussions in mind, the overarching objective of this paper is to optimize and schedule tasks by offloading the data from a busy \textit{primary node} to a relatively idle \textit{auxiliary node}. To address this objective, we propose the following contribution. 

\noindent\textbf{ (i) Data-Driven Resource-Aware Offloading Framework.} This framework optimizes the system parameters, such as the processing complexity of the task, memory utilization, bandwidth, and power availability, to assert the \textit{primary node} to offload a portion of the data to an \textit{auxiliary node}. Besides, we have introduced a novel parameter called \textit{split-ratio}, which helps us efficiently offload the data to the \textit{auxiliary node}. Our analysis indicates that offloading with a \textit{split-ratio} of 0.7-0.8 enhances the performance of task execution at the expense of increased power and memory usage (\textit{primary} + \textit{auxiliary node}). 

\noindent\textbf{ (ii) Testbed Development for System Evaluation.} A system of two Nvidia Jetson devices and two UGVs was designed to evaluate the optimization framework. The devices were equipped with an MQTT-based publisher-subscriber protocol to share the \textit{auxiliary node's} system parameters to the \textit{primary node} and offload the data to the \textit{auxiliary node}. To show the effectiveness of our proposed optimization framework, we assess its performance on DNN applications with multiple data modalities, including posture estimation (identifying human postures like standing, sitting, or lying down), semantic segmentation (classifying each pixel in an image to the object it belongs to, providing a detailed understanding of the scene), and object detection. 

\noindent\textbf{ (iii) Data Compression for Enhanced Optimization Performance.} As the data size grows, offloading data becomes more expensive. As a result, a frame compression and masking technique was leveraged to eliminate similar frames and extract the object of interest from the data to be offloaded, thereby reducing inference time and communication overhead, eventually enhancing overall performance in data-intensive environments.


\noindent\textbf{ (iv) Simulating Real-World Scenarios to Evaluate Offloading Strategy.} In real-world autonomous system operations, the individual nodes would be in motion, suggesting that the distance between the nodes can affect the network parameters. As a result, we simulate a scenario where the nodes are constantly in motion. As the distance increases, offloading latency rises, prompting questions about when to stop offloading images. This exploration offers valuable insights such as understanding the distance-offloading-latency relationship, identifying offloading thresholds for efficiency, and optimizing image offloading strategies in dynamic environments, ultimately improving performance in real world scenarios.

\section{\textbf{Related work}}\label{sec:related_work}

{\bf Edge inference and Offloading of multiple concurrent DNN tasks.} Inferencing Multiple DNN tasks concurrently is a critical salient feature for any autonomous system's real-time operation. Motivated by this, the authors of heimdall~\cite{Heimdel} developed a mobile GPU coordination platform for emerging augmented reality applications in which frame rates decrease and inference latency increases significantly due to multi-DNN GPU contention. It is designed with a pseudo-preemption mechanism that (i) breaks down the multi-DNN into smaller units and (ii) prioritizes and flexibly schedules simultaneous GPU tasks. Additionally, in BAND\cite{jeong2022band}, the authors develop a mobile-based inference system. BAND dynamically enables the creation of DNN execution plans and schedules DNNs on processors according to stated scheduling goals. In contrast, we consider a broader range of edge devices and application scenarios in this work to eventually minimize bandwidth consumption and latency. 

{\bf Task-based Scheduling}: Task-based scheduling plays a vital role in optimizing and reducing inference with respect to the assigned tasks. The authors of~\cite{roy2010supporting} propose a novel approach to the constraint optimization problem and suggest a greedy heuristic approach to choosing the best subset of concurrent applications within the constrained fidelity and resource budget.  Additionally, LaLaRAND~\cite{kang2021lalarand}, a real-time layer-level DNN scheduling framework that enables CPU/GPU scheduling of individual DNN layers with fine-grained CPU/GPU allocation schemes. This work tackles the schedulability of real-time DNN tasks, the asymmetric nature of DNN task execution on CPU and GPU, respectively, and the lack of task-based scheduling of CPU/GPU-aware allocation schemes. In contrast, our work involves running multiple DNN applications concurrently by splitting data into different nodes, considering each device's capabilities and task requirements.


{\bf Frame-based compression techniques.} Prior research has shown that optical flow can be utilized to estimate object motion across multiple camera views, enabling the system to track objects moving between cameras~\cite{multiview}. This information helps schedule video frame processing, minimizing latency and prioritizing relevant frames to meet real-time video analytics demands. On the other hand, AdaMask \cite{Adamask} presents an adaptive frame masking approach for efficient video streaming and processing in edge computing environments, focusing on lower communication overhead and accelerated DNN inference. However, both~\cite{multiview} and \cite{Adamask} aim at compressing frames acquired from static cameras. In contrast, we consider both static and mobile autonomous devices to optimize image offloading strategies, enhancing efficiency and effectiveness in various real-world scenarios.

\section{System Overview}
\label{sec:sysoverview}
\begin{figure}[htbp]
    \centering
    \includegraphics[scale=0.55]{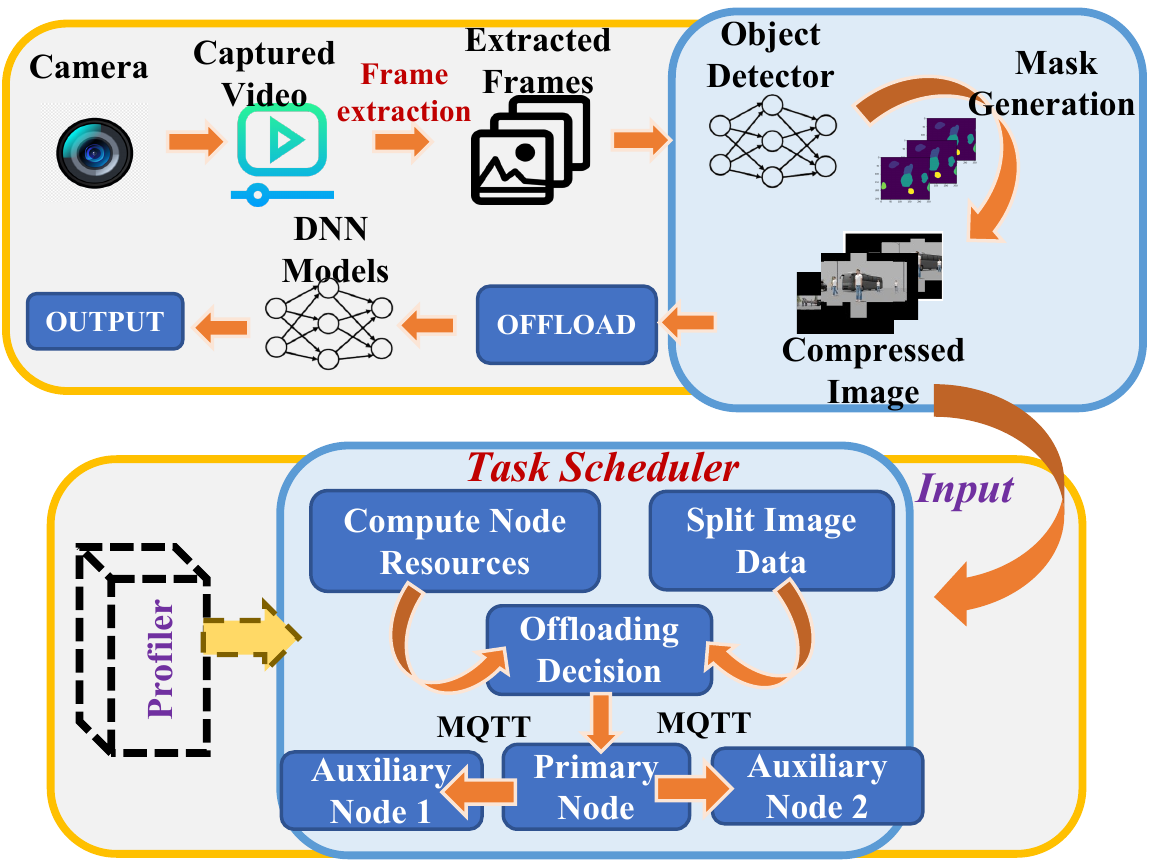}
    \caption{Optimization framework overview: Frame compression technique and task scheduling process.}
   \label{fig:framework}
\end{figure}

Our system architecture comprises a Device profiler and an online scheduler as shown in Fig. \ref{fig:framework}. The inputs are the image data and it is split according to resource availability.

We consider three important parts to design the framework \ref{fig:framework}. They are frame masking, profile engine, and optimization. This involves devising an efficient frame masking solution to ensure optimal performance during offloading between edge devices and creating a profiling engine that accurately evaluates primary and auxiliary nodes' performance, considering memory, power, and inference time while adapting to dynamic conditions like UGV movement. An optimization framework is also required to identify optimal split ratios for offloading decisions within specified bounds, resulting in a comprehensive and adaptive solution for multi-DNN systems.

\subsection{\name Components}

\textbf{Device profiler. }
In order to make memory and power-aware scheduling, we  analyze and monitoring of the performance of both devices, with a focus on key metrics such as device memory, power usage, inference time, and network latency. By gathering this information, we are able to evaluate the resource availability of each device in real-time and make informed decisions on the allocation of processing tasks.

\textbf{Task scheduler. }
In our optimization framework, we incorporated a task scheduler that intelligently manages task offloading in a multi-node environment. By gathering profiling data about primary and auxiliary devices, it assesses resource availability and the Multi-DNN workload. The task scheduler then ascertains if the primary node requires offloading and calculates the optimal data-split ratio for efficient resource utilization. Acting as a smart decision-making system,  It distributes workloads based on resource availability and performance capabilities, leading to reduced inference time, less memory utilization, and enhanced overall system efficiency.






The are some difficulties in designing an efficient offloading framework for multi-DNN execution systems in resource-constrained edge environments. This framework must address device heterogeneity, resource constraints, performance variability, and energy efficiency while adapting to dynamic conditions like varying velocities and distances. In moving conditions, the offloading latency may vary due to changes in distance between devices, leading to potential inefficiencies.
Processing large image data in edge environments poses challenges, as increased data transmission consumes more bandwidth and results in higher latency. Additionally, processing larger images requires more computational power, straining resource-constrained devices. To tackle this issue, we introduce a frame compression technique, specifically frame masking which is crucial for ensuring optimal performance during offloading. Frame masking reduces image data size, minimizing bandwidth consumption, lowering latency, and improving energy efficiency.

\subsection{System Assumptions}
In this context, we assume that resource-constrained devices have limited processing power and memory resources compared to more powerful servers or computers. When running multiple DNN models simultaneously, higher energy consumption and potential performance degradation are expected. Model inference may also be a concern when executing multiple DNN models concurrently on these devices. Additionally, scalability concerns arise due to the UGV's limited capacity to handle an increasing number of DNN models or growing complexity.
The system assumes a variety of UGVs with different processing capabilities, memory capacities, and energy consumption profiles. 
The UGVs are connected through a communication network that allows for data offloading and communication between devices. The profiling engine can accurately measure performance metrics like memory utilization, power consumption, and inference time. The system assumes that the UGVs may be in motion, causing the distance between them to change and affecting communication latency

\section{\textbf{\name Profiling Engine}}
\label{sec:profiling}
In our proposed system, the individual nodes continuously monitor system variables under multi-DNN workloads to identify optimal collaborative configurations. We first describe the testbed setup we used in profiling multi-DNN workloads across heterogeneous systems and then present quantitative insights from profiling various device and network attributes.

\subsection{Testbed setup}
\begin{figure}[htbp]
    \centering
    \includegraphics[width=0.5\textwidth]{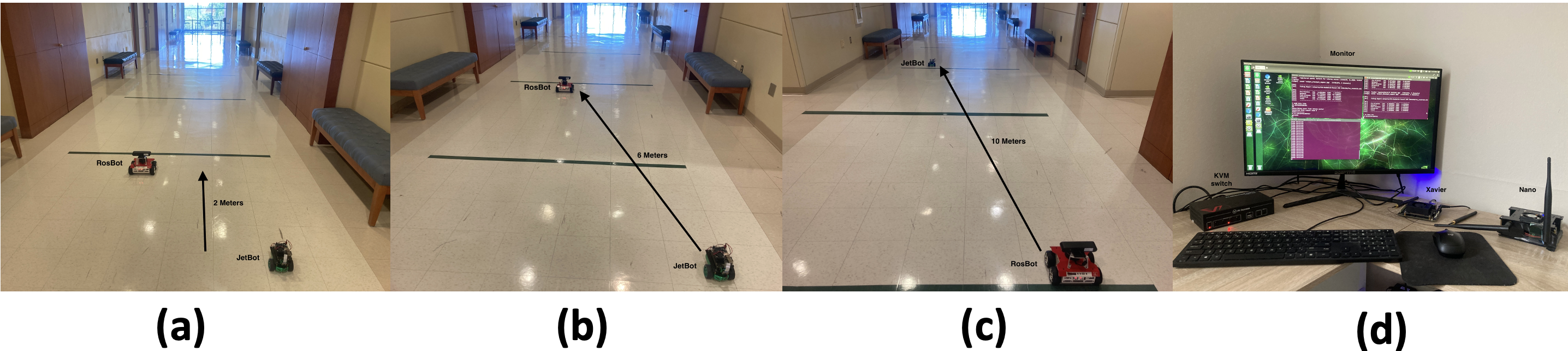}
    \caption{Experimental setup for Testbed: (a) UGV setup for 2-meter distance (b) 6-meter distance (c) 10-meter distance (d) static device (Nano-Xavier) setup.}  
    \label{fig:testbed_setup}
\end{figure}

We consider a network consisting of two heterogeneous edge platforms akin to a pair of autonomous systems with heterogeneous resources: (i) a low-resource Jetson Nano with a quad-core ARM Cortex-A57 MPCore processor, 4GB of LPDDR4 memory, and a 128-core NVIDIA Maxwell GPU, and (ii) a Jetson Xavier embedded with an octa-core NVIDIA Carmel ARM v8.2 CPU, 8GB LPDDR5, and a 512-core Volta GPU. In all our experiments, we assume that the lower end device (i.e., Nano) constantly monitors system parameters to offload its workload to the more powerful device, for executing multiple DNNs for downstream applications. Fig. \ref{fig:testbed_setup} shows the experimental setup of our testbed. While the Jetson Xavier device was positioned in a fixed location, UGVs mounted with Jetson Nanos were moved at different angles and velocities for emulating various mobility conditions. We adopted a publisher-subscriber architecture \cite{publisher_subcriber} (specifically, the Message Queuing Telemetry Transport (MQTT)~\cite{MQTT} protocol)  for message passing between the two devices.

\subsection{Device Profiling}

\name profiling engine runs on both primary and auxiliary nodes to continuously log memory utilization, power consumption, and inference time for both devices. While in our experiments we consider the low-resource device (i.e., Jetson Nano) as the primary node and Jetson Xavier as the auxiliary node for simplicity, in reality, all nodes in the network can assume primary and auxiliary roles. \name uses Jetson Stats~\cite{jetsonstats} to measure memory utilization and average power consumption. 

\textbf{DNN Workloads.} As \name is designed for autonmous systems that are required to run multiple concurrent compute-intensive tasks, we run two exemplar tasks, namely semantic segmentation and posture estimation,  using a multiprocessing pool. \name utilizes the Nvidia Jetson Inference Library~\cite{jetson_inference} for profiling the various DNN models in order to optimize offloading decisions.

\textbf{Split Ratio ($r$). }In our work, we propose the \textbf{split ratio}, which represents the proportion of images offloaded to the auxiliary node. It ranges from 0 (all images processed locally) to 1 (all images offloaded to the auxiliary). The optimal split ratio maximizes the collaborative system's throughput while minimizing resource consumption. Notations $T1$, $P1$, and $M1$ represent operation time, power, and memory usage of the auxiliary, while $T2$, $P2$, and $M2$ represent those of the primary node. $Offlatency$ refers to the network latency resulting from offloading images.

In Table~\ref{tab:profile_table}, we report the measured performance of the two devices in processing a batch of 100 images, under various configurations -- $r$ ranging from 0 to 1. As anticipated, we observe that while the overall power consumption is comparable between the nodes, the processing latency is significantly lower for the auxiliary device for the same workload. For e.g., at a $r=0.5$, while the processing time on the primary ($\approx$ 28.35 seconds) is double that of the auxiliary ($\approx$ 13.88 secs), the power consumption is comparable at 5.63 W and 5.42 W. At the same time, we also note that the offloading latency varies only minimally (between 0 and 1.56 secs) with $r$, supporting our premise for intelligent offloading.
\begin{table}[t]
\centering
\caption{Profiling results from testbed for semantic segmentation and posture estimation model}
\label{tab:profile_table}
\setlength{\tabcolsep}{4pt}
\renewcommand{\arraystretch}{1.2}
\resizebox{\columnwidth}{!}{%
\begin{tabular}{|c|c|c|c|c|c|c|c|c|}
\hline
{\color[HTML]{000000} \textbf{\begin{tabular}[c]{@{}c@{}}r \\ (split ratio)\end{tabular}}} &
  {\color[HTML]{000000} \textbf{\begin{tabular}[c]{@{}c@{}} \textbf{$T_{1}$} \\ (Xavier) \\ (s)\end{tabular}}} &
  {\color[HTML]{000000} \textbf{\begin{tabular}[c]{@{}c@{}} \textbf{$P_{1}$} \\ (Xavier)\\ (w)\end{tabular}}} &
  {\color[HTML]{000000} \textbf{\begin{tabular}[c]{@{}c@{}} \textbf{$M_{1}$} \\  (Xavier)\\ (\%)\end{tabular}}} &
  {\color[HTML]{000000} \textbf{1-r}} &
  {\color[HTML]{000000} \textbf{\begin{tabular}[c]{@{}c@{}} \textbf{$T_{2}$} \\  (Nano)\\ (s)\end{tabular}}} &
  {\color[HTML]{000000} \textbf{\begin{tabular}[c]{@{}c@{}} \textbf{$T_{3}$} \\ (Offlatency)\\  (s)\end{tabular}}} &
  {\color[HTML]{000000} \textbf{\begin{tabular}[c]{@{}c@{}} \textbf{$P_{2}$} \\ (nano) \\ (w)\end{tabular}}} &
  \textbf{\begin{tabular}[c]{@{}c@{}} \textbf{$M_{2}$} \\ (nano)\\ (\%)\end{tabular}} \\ \hline
{\color[HTML]{000000} 0} &
  {\color[HTML]{000000} 0} &
  {\color[HTML]{000000} 0.95} &
  {\color[HTML]{000000} 10.2} &
  {\color[HTML]{000000} 1} &
  {\color[HTML]{000000} 68.34} &
  {\color[HTML]{000000} 0} &
  {\color[HTML]{000000} 5.89} &
  {\color[HTML]{000000} 69.82} \\ \hline
{\color[HTML]{000000} 0.3} &
  {\color[HTML]{000000} 8.45} &
  {\color[HTML]{000000} 4.59} &
  {\color[HTML]{000000} 36.67} &
  {\color[HTML]{000000} 0.7} &
  {\color[HTML]{000000} 39.03} &
  {\color[HTML]{000000} 0.43} &
  {\color[HTML]{000000} 5.35} &
  {\color[HTML]{000000} 63.77} \\ \hline
{\color[HTML]{000000} 0.5} &
  {\color[HTML]{000000} 13.88} &
  {\color[HTML]{000000} 5.42} &
  {\color[HTML]{000000} 45.61} &
  {\color[HTML]{000000} 0.5} &
  {\color[HTML]{000000} 28.35} &
  {\color[HTML]{000000} 0.89} &
  {\color[HTML]{000000} 5.63} &
  {\color[HTML]{000000} 52.54} \\ \hline
{\color[HTML]{000000} 0.7} &
  {\color[HTML]{000000} 16.64} &
  {\color[HTML]{000000} 5.73} &
  {\color[HTML]{000000} 51.23} &
  {\color[HTML]{000000} 0.3} &
  {\color[HTML]{000000} 19.54} &
  {\color[HTML]{000000} 1.25} &
  {\color[HTML]{000000} 4.75} &
  {\color[HTML]{000000} 45.58} \\ \hline
{\color[HTML]{000000} 0.8} &
  {\color[HTML]{000000} 17.24} &
  {\color[HTML]{000000} 6.17} &
  {\color[HTML]{000000} 56.96} &
  {\color[HTML]{000000} 0.2} &
  {\color[HTML]{000000} 13.34} &
  {\color[HTML]{000000} 1.44} &
  {\color[HTML]{000000} 4.48} &
  {\color[HTML]{000000} 40.34} \\ \hline
{\color[HTML]{000000} 1} &
  {\color[HTML]{000000} 19.001} &
  {\color[HTML]{000000} 6.38} &
  {\color[HTML]{000000} 59.37} &
  {\color[HTML]{000000} 0} &
  {\color[HTML]{000000} 0} &
  {\color[HTML]{000000} 1.56} &
  {\color[HTML]{000000} 0.77} &
  {\color[HTML]{000000} 16} \\ \hline
\end{tabular}%
}
\end{table}

\subsection{Network Profiling}
\begin{figure}[thb]
    \centering
    \includegraphics[height=3.2cm, width=.5\textwidth]{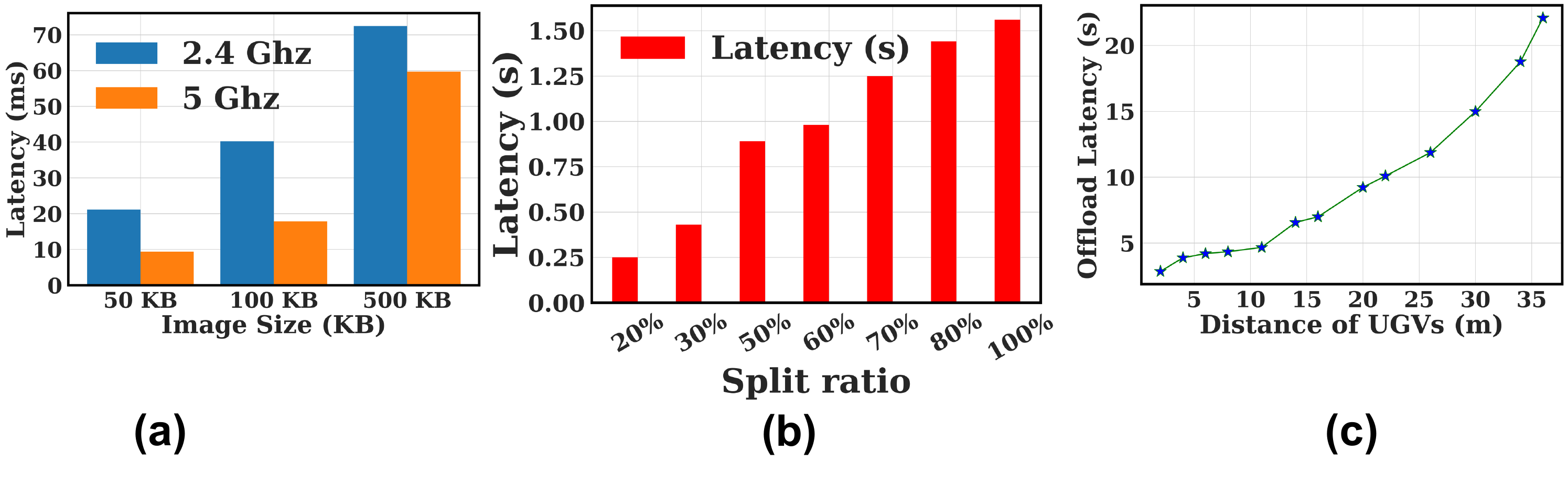}
    \hfill
    \caption{MQTT latency for (a) different network bands \& image sizes,  (b) different split ratios, \& (c) different distances with differing velocities of UGVs.}
    \label{fig:mqtt_latency}
\end{figure}

Furthermore, We investigated the network latency under two different network configurations- specifically, WiFi on two frequency bands, 2.4 Ghz and 5 Ghz. In Fig. \ref{fig:mqtt_latency}, we plot the latency ($y-$axis) (a) for different sizes of images, (b) various split ratios, and (c) distance between the primary and auxiliary devices, on the $x-$axis. We note that the higher band offers lower latencies and we observe increasing latencies with both increasing split ratios as well as distances.


In the next section, we describe the \name solver which takes into account the profiled variables to output an optimal collaborative system for vision tasks.
\section{\textbf{\name Solver}}
\label{sec:method}
We design the \name solver such that it dynamically adjusts data-splitting ratios based on available resources, optimizing the collective throughput and resource utilization. This cost-effective and scalable solution is applicable to various edge computing scenarios, addressing resource limitations and energy constraints, while allowing efficient processing of large data volumes in a distributed manner. In the following subsections, we describe the steps in devising the optimization framework.

\subsection{Latency and Energy Modeling}
In Table~\ref{tab:notation_table}, we list the variables used in the optimization formulation.

\begin{table}[htbp]
\caption{Important Notation \& Meaning.}
\label{tab:notation_table}
\centering
\small
\setlength{\tabcolsep}{4pt}
\renewcommand{\arraystretch}{0.4}
\begin{tabular}{|l|l|}
\hline
\multicolumn{1}{|c|}{Notation} & \multicolumn{1}{c|}{Meaning}                                                                                 \\ \hline
I                              & Input size of computational task                                                                             \\ \hline
N                              & \begin{tabular}[c]{@{}l@{}}Number of cpu cycle needed to execute \\ one bit of computation data\end{tabular} \\ \hline
$C_{cpu}$                      & Total cycle needed to finish specific task                                                                   \\ \hline
$T_{exec}$                     & Total execution time                                                                                         \\ \hline
S                              & Computation speed of a client device                                                                         \\ \hline
$E_{exec}$                     & Total execution energy                                                                                       \\ \hline
B                              & Transmission Bandwidth                                                                                       \\ \hline
$D_{R}$                        & Data rate                                                                                                    \\ \hline
$T_{0}$                        & Offloading latency                                        \\ 
\hline
$r$                        & Split ratio     \\                                                         
\hline
$T_{s}$                        & Time required to run ratio offloading code                                                                   \\ \hline
$E_{0}$                        & Offloading energy                                                                                            \\ \hline
$E_{s}$                        & Energy required to run ratio offloading code                                                                 \\ \hline
\end{tabular}
\end{table}


\subsubsection{Execution Period}
$I$ denotes the input size of the computation task of the offloading device. N presents the number of CPU cycles needed to execute one bit of input computation data. Therefore, $C_{cpu}=NI$ denotes the cycles needed to finish the selected computation task. So, the executing latency can be expressed as $T_{exec}=\frac{C_{cpu}}{S}$ where $S$ is the computation speed of the device and it is measured in cycles per second. 

We model the power consumption of CPU as $P=\mu(S)^{3}$ as in \cite{zhang2013energy} Thus, the energy consumption per cycle is $\mu(S)^{2}$. The energy consumption for deep model processing is then $E_{exec}=C_{cpu}\mu(S)^{2}$ where $\mu$ is a coefficient depending on chip architecture. We assume that the computing speed of the server's CPU is limited to $S^{max}$, so we may have $0\leq S \leq S^{max}$. Considering $r$ is the split ratio,
\begin{equation*} E_{exec}=E_{1}r + E_{2} (1-r) \end{equation*}
\begin{equation*} T_{exec}=T_{1}r + T_{2} (1-r) \end{equation*}
Here, $E_{1}$ and $E_{2}$ are the execution energy of the two nodes participating to complete one task. Similarly, $T_{1}$ and $T_{2}$ are the execution time of the two devices for a specific split ratio.
\subsubsection {Offloading Period}
If $B$ is transmission bandwidth, $d$ is distance between two devices, $e$ is path loss exponent, $N_{0}$ is the Gaussian noise power, the transmission data rate for offloading task can be found using the Shannon-Hartley algorithm \cite{Shannon}.
\begin{equation*} D_{R}=B\ \log_{2}(1+\frac{d^{-u}P_{t}}{N_{0}}) \end{equation*}
Here, $P_{t}$ is the transmission power of the device during offloading. If the medium is lossless then we can put $u=0$.

Then, the Offloading Latency is $T_{o}= \frac{C}{D_{R}}$. Here, $C$ will depend on the selected split ratio.

The Total latency can be given by $T=T_{exec}+T_{o}+T_{s}$. Energy required to run the split ratio selection code is then $E_{s}=P_{k}T_{s}$ where $P_{k}$ is the power rating of the device which will run the solver code.

Offloading energy requirement can be divided into two parts, 
\begin{equation*} 
E_{0}=T_{0}\sum_{i=0}^{N}P_{i}
\end{equation*}
Here, $P_{t}$ is the power required of the sender device and $P_{r}$ is the power drawn while receiving the sent data. Then the Total energy can be expressed as $E=E_{exec}+E_{s}+E_{o}$.

\subsubsection{Solver}
As our objective is to make the system both memory and energy-aware while minimizing the latency (i.e., improving the overall throughput), we have derived a relation between energy and memory during execution time. Specifically, we have considered the quadratic relation between energy and required memory during execution. While solving for optimization, we can use the variable substitution approach. We can work with the real part only of the following equation for a sub-optimal solution.
\begin{equation*}
T=r(T_{1}+T_{3})+(1-r)T_{2}
\label{T}
\end{equation*}
Here T1 is the operation time for Jetson Xavier and $T_{2}$ is the operation time for Jetson Nano. $T_{3}$ is the round trip time for image transfer.
\begin{equation} T_{1} = a_{1}r^{2} + a_{2}r + c_{1}
\label{T1}
\end{equation}
\begin{equation*} T_{2}= b_{1}(1-r)^{2} + b_{2}(1-r) + c_{2}
\label{T2}
\end{equation*}
\begin{equation} E_{1} = a_{1}r^{3} +a_{2}r^{2}+ a_{3}r + c_{1}
\label{e1}
\end{equation}
\begin{equation*} E_{2} = b_{1}(1-r)^{3}+b_{2}(1-r)^{2} + b_{3}r + c_{2}
\label{e2}
\end{equation*}
\begin{equation} M_1 = a_{1}r^{2} + a_{2}r + c_{1}
\label{m1}
\end{equation}
\begin{equation*} M_2 = b_{1}(1-r)^{2} + b_{2}(1-r) + c_{2}
\label{m2}
\end{equation*}

Here, $E1$, and $E2$ are energy consumption for Jetson Xavier and Nano. $M1$ and $M2$ are also memory needed for Jetson Xavier and Nano
the values of $a_{1}$, $a_{2}$, $a_{3}$, $b_{1}$, $b_{2}$, $b_{3} $ coefficients can be found through curve fitting feature with some experimental values.
Problem formulation,
\begin{align}
\min (T) \quad &\mathrm{constraints:} \nonumber \\
&C1: T\leq \frac{\tau}{k}, \qquad &&C2: 0\leq P_k\leq P^{\max}, \nonumber\\ &C3: 0<r<1, \qquad &&C4: 0\leq S\leq S^{\max}, \nonumber\\&C5: E_{\mathrm{exe}}^k\leq W^k, \qquad &&C6: M_{\mathrm{exe}}^k\leq M^k.
\label{eq14} 
\end{align}

Here, $\tau$ is the execution latency while the whole operation is done by one device and $k$ is the total number of devices. $W^{k}$ and $M^{k}$ are the highest power and memory ratings of the individual devices respectively.

\subsubsection{Battery and Charging Constraints}
We pose additional constraints based on the remaining battery life (before a recharge is required), to specifically cater for our mobile autonomous systems.
To account for this, we consider two UGVs RosBot\footnote{\url{https://husarion.com/manuals/rosbot/}} and JetBot\footnote{\url{https://jetbot.org/master/}} with a 4000 mAh battery capacity and a discharge rate of 70\%. These run on computing platforms similar to those considered in our previous analyses. The UGVs have a driving time of about 20-25 minutes and lose around 15-20 watts during that time. Running an additional DNN model on the bot will reduce its power capacity even further. If our DNN model runs for 50-60 seconds, it will consistently consume 5 to 6 watts of power. Equations \ref{e18} and \ref{e19} show how we calculate available power given these additional constraints. 


The optimization is then based on an additional power threshold constraint, and when the available power (P) falls below the threshold, the UGV starts offloading more aggressively to the auxiliary system (recall that the power consumption of running concurrent DNNs is more than that of offloading images, see Table~\ref{tab:profile_table}). This ensures that the UGV can complete its mission without running out of power.

\begin{equation}
E_{available}=C_{0}k-E_{dnn}-E_{drive}
\label{e18}
\end{equation}
\begin{equation}
P_{available}=E_{available}/\left ( \left ( 1-k \right )\left ( t_{dnn}+t_{drive} \right )/3600 \right )
\label{e19}
\end{equation}

Here, $E_{available}$ represents the available energy, $k$ is the discharge rate of the battery, while $E_{dnn}$ and $E_{drive}$ denote the energy consumption of running the DNN model and driving the UGV, respectively.
$t_{dnn}$ is the duration of the DNN model in seconds,
$t_{drive}$ is the duration of the UGV driving in seconds. Using curve fitting, we derive memory and power as functions of inference time. We found empirically that quadratic functions have the best fit with adjusted $R^2$ values of 0.976 and 0.989, respectively.

\subsubsection{Mobility Constraints}
In order to address the offloading latency in dynamic scenarios where , we introduce a minimum threshold for offloading latency, denoted as $\beta$. This threshold helps to manage the offloading process more efficiently by considering the changing motion of the UGVs.

To calculate the distance between the UGVs, we employ the following equation:
\[
d = (V_{\text{primary}} + V_{\text{auxiliary}}) \times t
\]
This equation calculates the distance, $d$, between two UGVs based on their velocities, $V_{\text{primary}}$ and $V_{\text{auxiliary}}$, and a given time interval, $t$. The equation takes into account the relative motion of the UGVs, and the distance increases as the UGVs move apart during the time interval.
The relationship between latency (L) and distance (d) between the two UGVs is modeled using the following equation and we get this equation from curve fitting:
\[
L = a_{1} \times d^2 - a_{2} \times d + a_{3}
\]
This equation represents the time delay (latency) in sending images from one UGV to another. As the distance, $d$, between the devices increases, the latency, $L$, also increases, affecting the efficiency of the offloading process.

When the latency meets or exceeds the threshold, $\beta$, the system stops sending data:
\[
\text{If } L \geq \beta, \text{ stop sending data}
\]
This approach ensures that the offloading process is adapted to the dynamic changes in UGV motion, resulting in more efficient resource utilization and improved overall performance.

\subsection{Implementation}
We use Python's GEKKO library\cite{Gekko} to solve our optimization problems. The objective function, variables and constraints are all specified as we described earlier with a nonlinear optimization solver (i.e., IPOPT solver \cite{IPOPT}).


\begin{algorithm}
	\caption{Algorithm for Split Ratio Selection}
	\begin{algorithmic}[1]
	\Require IPs of connected devices $n$, Memory profiling of the nodes $M$, Inference time of device 1 $T_1$, Inference time of device 2 $T_2$, Round trip time $T_3$
	\Ensure Determine the split ratio $r$ for optimal operation time
            \State On the primary node:
            \State Calculate the device availability factor $\lambda$ based on the memory of both devices.
	    \textbf {Compute} the coefficients $a_{1}$, $a_{2}$, $b_{1}$, $b_{2}$, $c_{1}$, $c_{2}$ from equations~\ref{m1} using curve fitting 
           \State \textbf {if} {$M_{1},M_{2} \geq \lambda$} and  check latency, $L\leq \beta$ \textbf{then}
                \State Assign constraints from equation 3 on the following objective:
                    \begin{equation*} 
                    T = r(T_1 +T_3) + (1-r) T_2
                    \end{equation*}    
             \State Check battery capacity and available UGV power: \ref{e18} and \ref{e19}
                    \begin{equation*} 
                   P_{available}=E_{available}/\left ( \left ( 1-k \right )\left ( t_{dnn}+t_{drive} \right )/3600 \right )
                    \end{equation*}  
                      if $P_{available}>=E_{max}$
	    \State Solve the formulated problem for the given constraints using Interior Point Optimizer method
	    \State Send the derived amount of data to the subscriber node
	\end{algorithmic} 
\end{algorithm}

\section {\textbf{Frame-level Compression}}
\label{sec:compression}
\begin{figure}[htbp]
    \centering
    \includegraphics[height=2.3cm,width=0.5\textwidth]{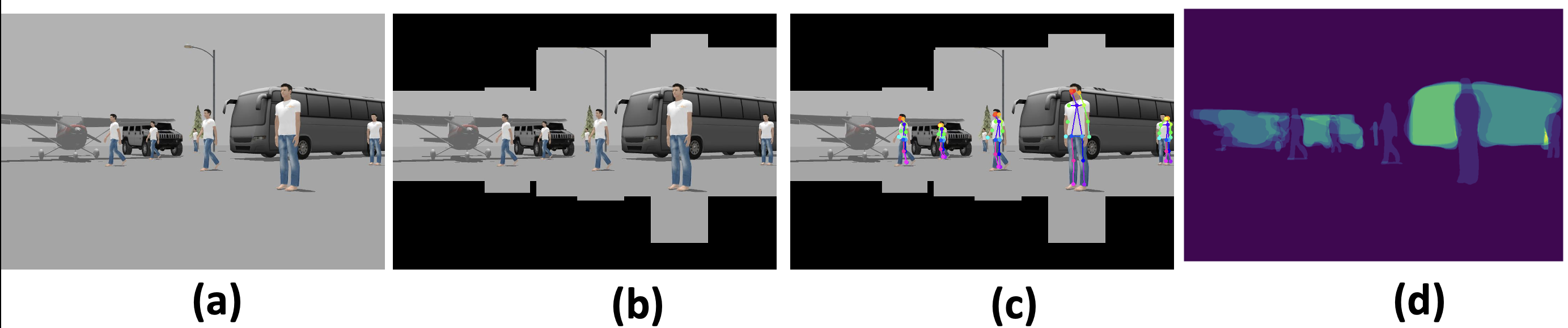}
    \caption{(a) Original frame (b) Compressed frame \& input  (c) Results from compressed frame for Pose Estimation  (d) Results from compressed frame for Semantic Segmentation.}
   \label{fig:original_vs_masked}
\end{figure}

To further optimize performance, \name teases out \emph{regions of interest} in images prior to running downstream DNN inferences (e.g., pose detection, image segmentation, etc.). This step ensures that the network latency from offloading them to an auxiliary node, when feasible, is also lowered. \name first uses a state-of-the-art object detection model that generates binary masks where pixels with \emph{detected objects} are denoted by bit 1, and 0 elsewhere. Element-wise multiplication of the binary mask with the original image returns a \emph{compressed image} (see Fig. \ref{fig:original_vs_masked}) which isolates objects of interest and eliminate extraneous backgrounds. 

To demonstrate the savings in bandwidth utilization and inference times, we conduct microbenchmark experiments using two downstream DNN models. We generated a virtual environment in the Gazebo simulator~\cite{Gazebo} generating 3100 images with a total of 9 common object classes such as persons and vehicles present. First, we use the faster-RCNN object detector~\cite{girshick2015fast} for generating compressed images which are then fed to exemplar downstream DNNs; semantic segmentation (SegNet~\cite{segnet}) and posture detection (PoseNet~\cite{posenet}) models, respectively. Figure~\ref{fig:original_vs_masked} shows illustrative examples of the output on the compressed frames for the two tasks. Overall, we observe a 13\% reduction (on a Jetson Nano device) in the total computational time corresponding to a savings in bandwidth by up to 28\% (i.e., from 8 MB down to 5.8 MB through compression) if the images were to be offloaded. While we observe only a 2\% drop in inference accuracy from compression, the astute reader will note that an imperfect object detector model may create artifacts for downstream computer vision tasks; we defer an in-depth study as future work.

\section {\textbf{Evaluation}}
This section presents our in-depth evaluation of the proposed optimization framework on the Gazebo-based dataset we describe in Section~\ref{sec:compression}, previously.


\subsection{Deriving constraints for optimization}

We derive constraints for optimization presented in equation \ref{eq14} which provides the limitations and requirements of different resources that must be satisfied during the optimization process. We consider deriving constraints for important resources i.e., memory, power and inference time for optimizing the task offloading process.
\begin{table*}[htbp]
\centering
\caption{Results from the real-time system for static condition.}
\label{tab:testbedresult_table}
\begin{tabular}{|c|c|c|c|c|c|c|c|}
\hline
{\color[HTML]{000000} \textbf{r (split ratio)}} &
  {\color[HTML]{000000} \textbf{\begin{tabular}[c]{@{}c@{}}T3 (Offloading \\ latency ) (s)\end{tabular}}} &
  {\color[HTML]{000000} \textbf{\begin{tabular}[c]{@{}c@{}}P1 (Xavier) \\ (w)\end{tabular}}} &
  {\color[HTML]{000000} \textbf{\begin{tabular}[c]{@{}c@{}}M1 (Xavier)\\ (\%)\end{tabular}}} &
  {\color[HTML]{000000} \textbf{1-r}} &
  {\color[HTML]{000000} \textbf{\begin{tabular}[c]{@{}c@{}}T1+T2 \\ (s)\end{tabular}}} &
  {\color[HTML]{000000} \textbf{\begin{tabular}[c]{@{}c@{}}P2 (nano)\\ (w)\end{tabular}}} &
  {\color[HTML]{000000} \textbf{\begin{tabular}[c]{@{}c@{}}M2 (nano)\\ (\%)\end{tabular}}} \\ \hline
0.2  & .67  & 4.87 & 32.09 & 0.8  & 55.38 & 6.96 & 75.12 \\ \hline
0.35 & 1.23 & 5.12 & 41.56 & 0.65 & 51.89 & 6.11 & 70.17 \\ \hline
0.45 & 1.98 & 5.78 & 49.55 & 0.55 & 42.87 & 6.24 & 65.66 \\ \hline
0.5  & 2.34 & 5.57 & 50.09 & 0.5  & 43.09 & 5.69 & 54.65 \\ \hline
0.6  & 2.90 & 6.35 & 53    & 0.4  & 39.45 & 5.88 & 57.77 \\ \hline
0.7  & 3.23 & 6.03 & 59.56 & 0.3  & 36.43 & 5.17 & 47.13 \\ \hline
0.8  & 3.55 & 6.34 & 63.45 & 0.2  & 34.90 & 5.35 & 43.34 \\ \hline
0.9  & 3.56 & 7.12 & 69.09 & 0.1  & 28.23 & 4.89 & 40.11 \\ \hline
\end{tabular}%
\end{table*}

\begin{figure}[http]
    \centering
    \includegraphics[width=.5\textwidth]{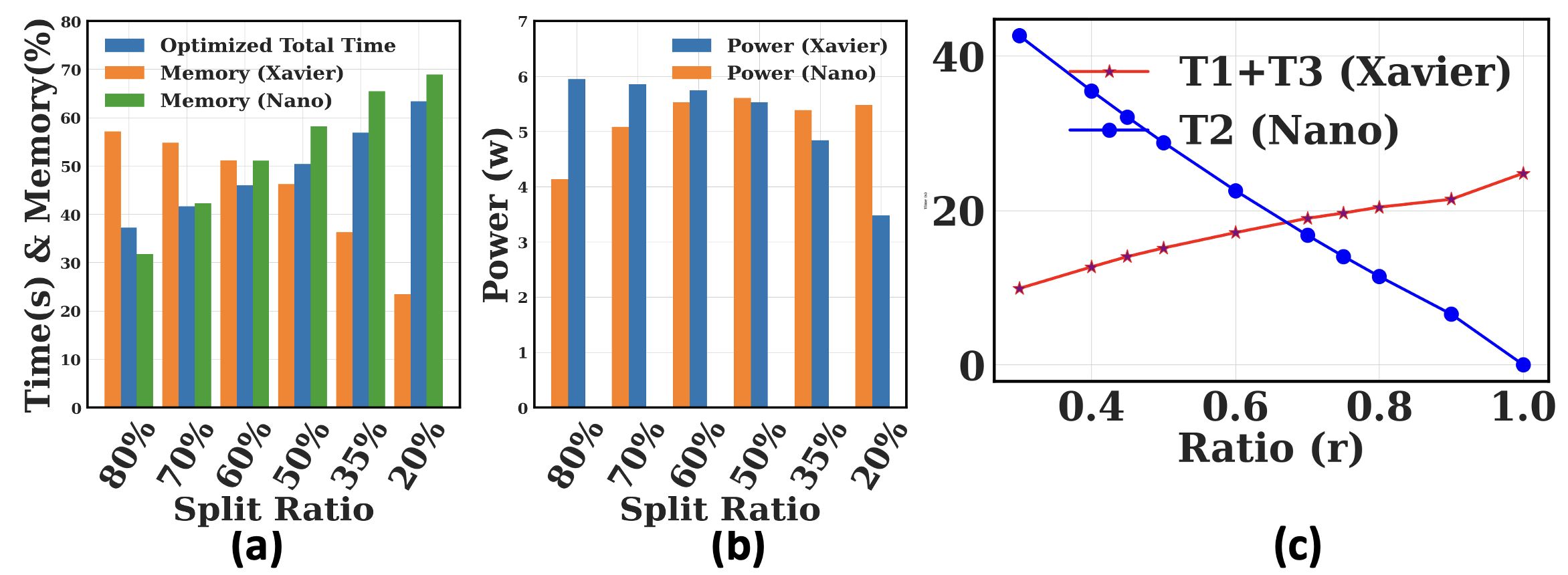}
    \hfill
    \caption{Optimized results for different split ratios. (a) Total time and (b) memory usage for different split ratios and power usage for both devices, (c) Computational time for both devices changes with split ratio r.}
    \label{fig:optimized_results1_4}
\end{figure}

From the experiments performed on the primary node, we obtain the base processing time for running multiple models on a single device. The total processing time for 100 images was found to be 68.34 seconds, as shown in Table \ref{tab:profile_table}. We got 200 output images from two models for 100 image inputs. We use this as baseline for inference time, where optimized inference time must be less than 68.34 seconds overall. As we described in Section~\ref{sec:method}, we use curve fitting to analyze the relationship between inference time and data splitting ratio, memory and splitting ratio, and power and splitting ratio which represent equations \ref{T1}\ref{e1}\ref{m1}.  This enables us to predict the inference time for different splitting ratios and allows us to 
identify the optimal data-splitting ratio that will provide the lowest inference time while considering memory and power constraints for the task offloading process.
Fig. \ref{fig:optimized_results1_4}(a) and Fig. \ref{fig:optimized_results1_4}(b) show the time, memory and power for different split ratios by our proposed \names~ solver .

From the solver we got the best value of the split ratio is 70\% within our desired memory and power constraints. The total inference time for this split ratio is 17.72 seconds for Xavier (70 images) and 16.79 seconds for Nano (30 images). On average for two models and 200 outcomes, it takes a total of 34.51 seconds. 

\subsection{Empirical Evaluation}
We perform empirical evaluations of static and dynamic conditions for our scenarios. We discuss the static and dynamic conditions in two case studies in the following.

\textbf{Case-1:} In this case, two UGVs are positioned at a fixed distance from each other, and their velocities are the same, leading to no relative movement between them. Since they are only 4 meters apart in our experiment, the communication overhead and latency remain constant throughout the evaluation. With a stable MQTT communication, the offloading of images from the primary UGV to the auxiliary UGV experiences a consistent latency based on their fixed distance. This constant latency allows an effective offloading process without additional communication overhead due to varying distances. The optimization framework can then be evaluated under this static condition to understand its performance in a controlled environment with minimal variations in latency.

The performance metrics for different ratios tested on the real-time testbed with the static condition are consistent with the optimization results from the solver, as illustrated in Table \ref{tab:testbedresult_table}. Here $T_{1}+T_{2}$ represents the total operation time for both Jetson Nano and Xavier and $T_{3}$ stands for the offloading latency for UGVs with the static condition of 4 meters distance from each other. From the results, we notice there is a slight change in offloading latency with split ratios. In this case, offloading latency increases only with higher split ratios and the distance between UGVs is constant.

After using the IPOPT optimization solver, we estimate that offloading 70\% of the images to a more powerful Xavier is the best optimal option given our specific memory and power constraints. We then evaluated the results in real-time systems (Table \ref{tab:testbedresult_table}) and tested different split ratios on a testbed and ended up with our desired results, which support our optimization framework. Overall, this demonstrates the efficacy of our framework in memory and power-aware task offloading, as well as the potential for further optimizations.

\textbf{Case-2:} In this case, the two UGVs are in motion with different velocities and/or directions, resulting in a dynamic distance between them. As the distance between the UGVs changes over time, the communication overhead and latency for offloading images from the primary to the auxiliary node also vary.
Due to the dynamic nature of this scenario, the MQTT communication may experience fluctuations in latency based on the changing distance between the UGVs. This can lead to challenges in the optimization framework, as the offloading process may need to account for variations in communication overhead and latency.
The optimization framework can be evaluated under this dynamic condition to understand its performance in a real-world environment with changing distances between the UGVs. This will help identify any potential issues and adjustments that may be needed to improve the framework's adaptability to varying communication conditions.\\
\begin{table*}[htbp]
\centering
\scriptsize
\caption{Overview of different deployed models in the testbed.}
\label{tab:different_model}
\begin{tabular}{|l|l|l|l|l|l|l|l|}
\hline
\multicolumn{1}{|c|}{\textbf{Application}} &
  \textbf{DNN Model} &
  \multicolumn{1}{c|}{\textbf{\begin{tabular}[c]{@{}c@{}}Split Ratio (r=0)\\ Run 100\% on\\ primary node \\ T2 (Nano) (s)\\ (Original image)\end{tabular}}} &
  \multicolumn{1}{c|}{\textbf{\begin{tabular}[c]{@{}c@{}}Split Ratio(r=0)\\ T2 (nano)\\ (s)\\ (Masked image)\end{tabular}}} &
  \multicolumn{1}{c|}{\textbf{\begin{tabular}[c]{@{}c@{}}Split Ratio (r=0.5)\\ T1 (Xavier)+\\  T2 (Nano)\\ (s)\\ (Original image)\end{tabular}}} &
  \multicolumn{1}{c|}{\textbf{\begin{tabular}[c]{@{}c@{}}Split Ratio (r=0.5) \\ T1 (Xavier)+ \\ T2 (Nano)\\  (s) \\ (Masked image)\end{tabular}}} &
  \multicolumn{1}{c|}{\textbf{\begin{tabular}[c]{@{}c@{}}Split Ratio (r=0.7)\\ T1(Xavier)+\\ T2 (Nano)\\ (s)\\ (Original image)\end{tabular}}} &
  \multicolumn{1}{c|}{\textbf{\begin{tabular}[c]{@{}c@{}}Split Ratio (r=0.7)\\  T1 (Xavier)+ \\ T2 (Nano) \\ (s) \\ (Masked image)\end{tabular}}} \\ \hline
\begin{tabular}[c]{@{}l@{}}Image recognition+\\ Object Detection\end{tabular} &
  \begin{tabular}[c]{@{}l@{}}ImageNet,\\  DetectNet\end{tabular} &
  74.68 &
  69.90 &
  56.74 &
  49.78 &
  44.13 &
  38.98 \\ \hline
\begin{tabular}[c]{@{}l@{}}Object detection+\\ Depth Sensing\end{tabular} &
  \begin{tabular}[c]{@{}l@{}}DetectNet,\\ DepthNet\end{tabular} &
  76.90 &
  71.34 &
  64.2 &
  57.89 &
  43.17 &
  40.32 \\ \hline
\begin{tabular}[c]{@{}l@{}}Semantic Segmentation\\ + Depth Sensing\end{tabular} &
  \begin{tabular}[c]{@{}l@{}}SegNet,\\ DepthNet\end{tabular} &
  71.25 &
  65.56 &
  58.43 &
  53.66 &
  48.37 &
  43.2 \\ \hline
\begin{tabular}[c]{@{}l@{}}Image recognition+\\ Depth Sensing\end{tabular} &
  \begin{tabular}[c]{@{}l@{}}ImageNet,\\ DepthNet\end{tabular} &
  69.66 &
  61.47 &
  50.64 &
  46.45 &
  43.54 &
  38.43 \\ \hline
\begin{tabular}[c]{@{}l@{}}Object Detection+ \\ Pose estimation\end{tabular} &
  \begin{tabular}[c]{@{}l@{}}DetectNet,\\ PoseNet\end{tabular} &
  67.28 &
  64.89 &
  51.59 &
  46.89 &
  39.69 &
  35.9 \\ \hline
\end{tabular}%
\end{table*}


\begin{figure*}[http]
    \centering
    \includegraphics[scale=0.23]{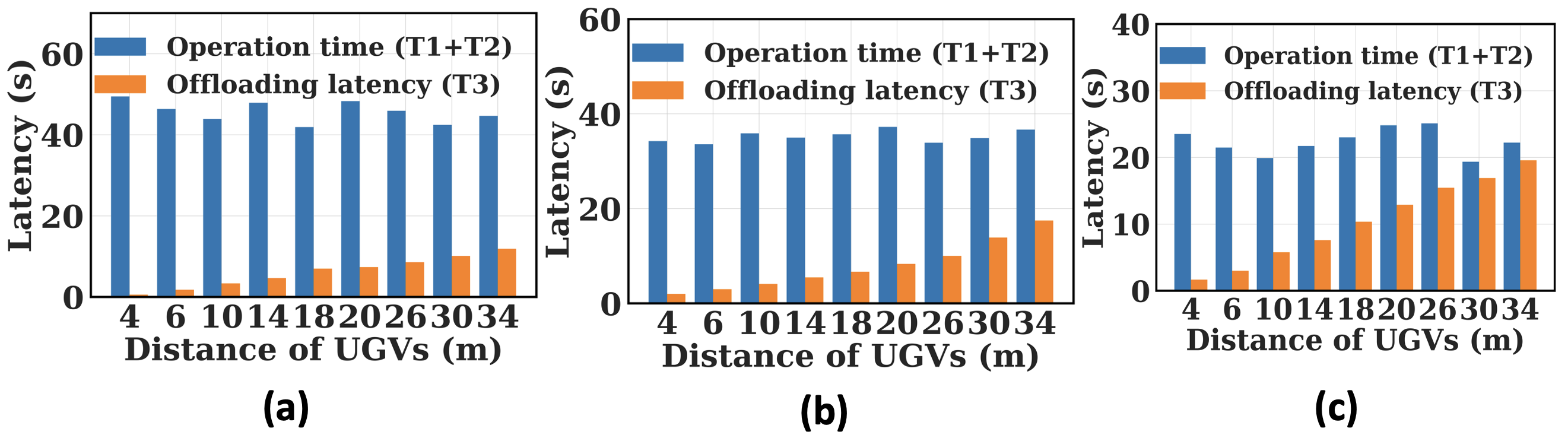}
    \hfill
    \caption{The experimental setup for the testbed is used to evaluate the performance of the system at different distances. (a) 30\% split ratio (b)70\% split ratio (c) 100\% split ratio.}
    \label{fig:evaluation_diff}
\end{figure*}

In  Fig. \ref{fig:evaluation_diff} we record total operation time ($T_{1}+T_{2}$) for both UGVs and offloading latency $T_{3}$ for the varying distances between UGVs. In this setup, we use three different split ratio 30\%, 70\% and worst case 100\% split and velocity for primary and auxiliary UGVs are respectively $V_{primary}$=1 m/s and $V_{auxiliary}$=3 m/s. The evaluation results reveal a positive correlation between the distance among the UGVs and the offloading latency. As the distance increases, the offloading latency also rises, affecting the optimization system's effectiveness. For instance, at a distance of 26 meters, the average offloading latency from the primary node to the auxiliary node is 13.9 seconds, which increases the communication overhead and compromises the system's performance.

In order to address this challenge, we propose an offloading latency threshold. The system can effectively track the offloading latency during the operation. If the latency surpasses the threshold, the primary node stops offloading images to the auxiliary node and searches for a more suitable split ratio lower than the previous one. If the search for an optimal split ratio within the bounds is unsuccessful, the primary node performs all processing tasks locally. This adaptive approach maintains optimal performance by minimizing communication overhead and avoiding excessive latency.


\subsection{Evaluation with Model Heterogeneity}

In order to thoroughly evaluate the performance of the {\em HeteroEdge}, it is critical to validate it with a diverse range of models that represent different use cases and applications. 
To accomplish this, we carefully selected a range of models that represent different types of computational requirements from object detection to image classification and depth estimation. As an exemplar, we selected computer vision model ImageNet\cite{imagenet} for object detection, DetectNet\cite{detectnet} for object localization and DepthNet\cite{depthnet} for monocular depth estimation. We deployed each model including the previous two models PoseNet and SegNet on our {\em HeteroEdge} testbed and ran them concurrently while measuring the key performance metrics such as total operation time, power consumption, and resource utilization. 

\begin{figure}[htbp]
    \centering
    \subfloat []{\includegraphics[width=0.23\textwidth]{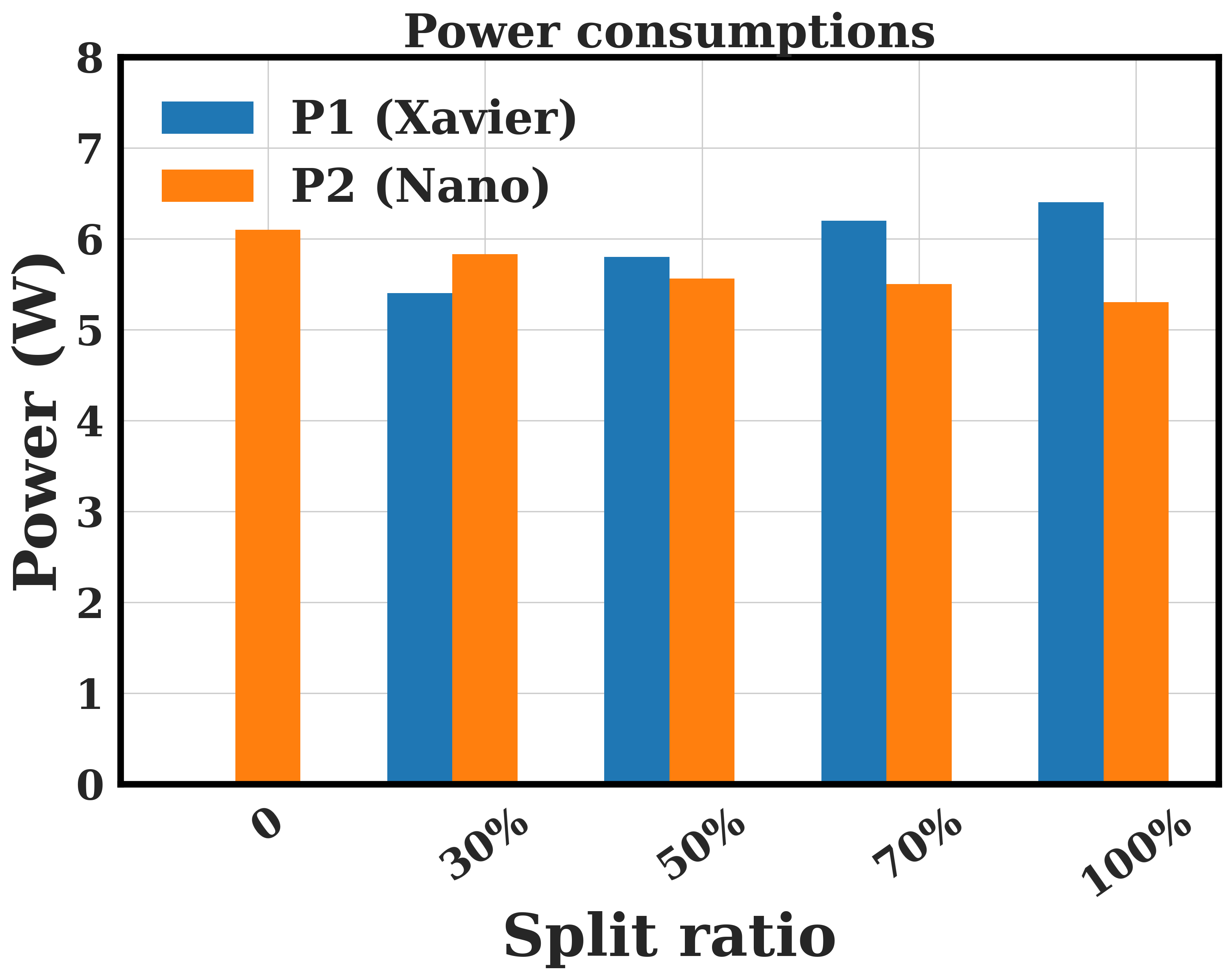}}
    \hspace{0.3cm}
    \subfloat []{\includegraphics[width=0.23\textwidth]{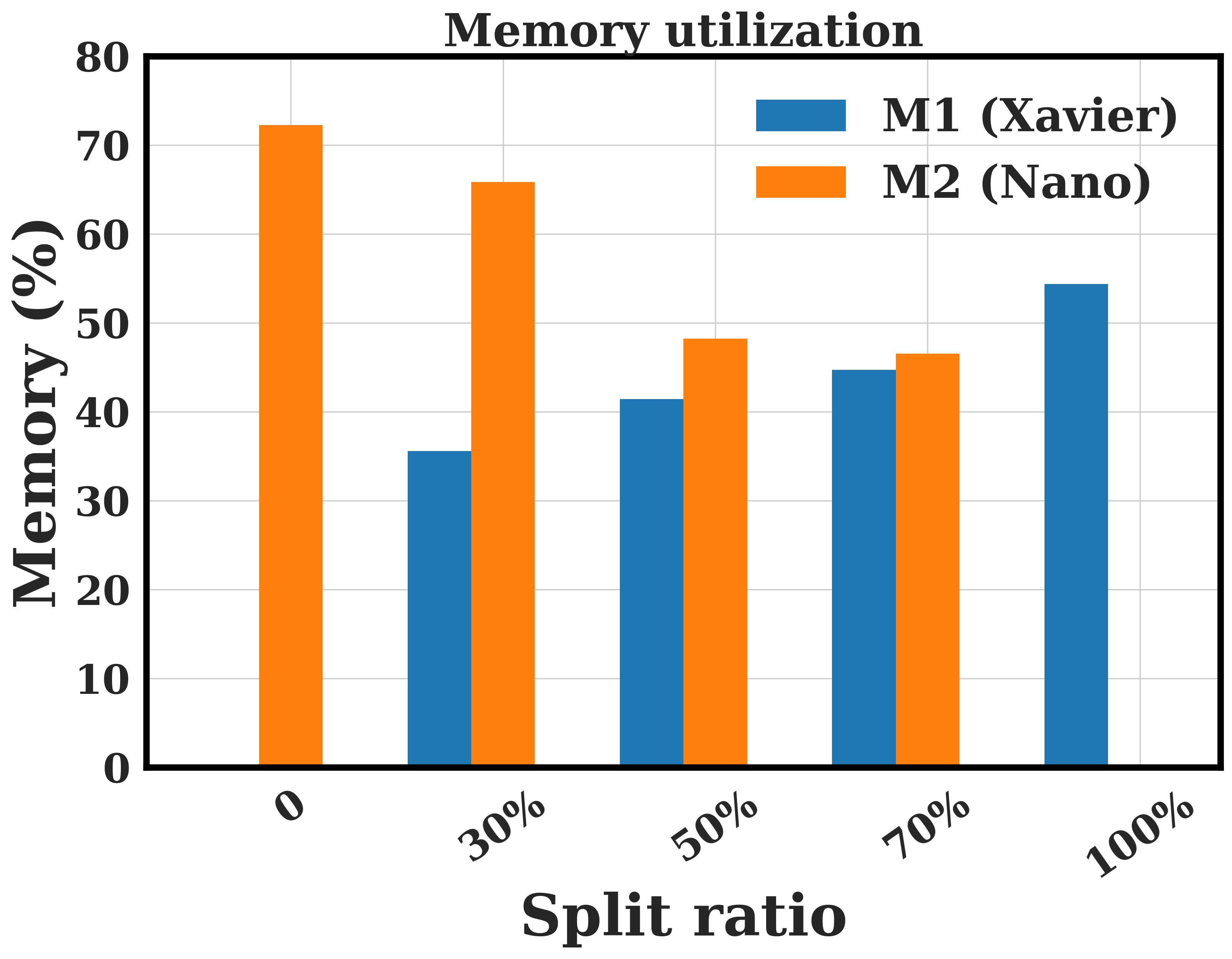}}
    \vspace{-1ex}
    \caption{(a) Average power consumption (b) Average memory utilization.}
    \label{fig:power_memory}
\end{figure}

Table \ref{tab:different_model} depicts the results from multiple DNN models simultaneously running on primary and auxiliary nodes. We tested it for 100 images and when the split ratio became zero, the entire processing was carried out only on the primary node as shown in Table \ref{tab:different_model}. We observe that the primary node sends 50\% and 70\% of the images to the auxiliary node for split ratio 0.5 and 0.7, respectively. It is worthwhile to note that there is an additional runtime on the primary node for object detection and mask generation. We record on average 3-4 ms latency per image with a lightweight faster-rCNN model~\cite{girshick2015fast}. We notice that the total operating time is lower (on average 9\%) in case of masked frames compared to the original frames though there was a notable change in power consumption and memory utilization. Fig. \ref{fig:power_memory}(a) presents a slight increase in power consumption which is on average 4-5\% more compared to the baseline where all processing is done locally for split ratio = 0. On contrary, Fig.\ref{fig:power_memory}(b) depicts a significant reduction in memory usage compared to the baseline memory usage i.e., $\approx$ 72.23\% at split ratio = 0. For example, for a 70\% split ratio, both devices use an average of 47\% of memory, which is almost a 34\% decrease compared to the baseline configuration. 

\section{\textbf{Conclusion \& Future work}}
In conclusion, our optimization work focused on reducing the latency of DNN model inference by offloading image processing to more powerful computing devices. We proposed framewise optimization a split ratio metric to determine the proportion of images to offload and used a solver to determine the optimal split ratio based on the memory and power constraints of UGV. Overall, our work provides a practical solution for reducing DNN model inference latency on resource-constrained devices. Our results show that offloading with MQTT and dynamic adjustment of the split ratio based on available power can further reduce latency and improve performance.
In this work, we utilize a primary-auxiliary node setting which is hierarchical offloading in nature. We want  to extend the current research to consider a star topology for offloading tasks in future work. In a star topology, a central node (the ``hub") manages the communication and coordination among multiple edge devices (the ``spokes"), allowing for more efficient resource allocation and data sharing.

\section*{\textbf{Acknowledgment}}
This work has been partially supported by NSF CAREER Award \#1750936 and U.S.Army Grant \#W911NF2120076.
\bibliographystyle{unsrt}
\bibliography{main}

\end{document}